\documentclass[letterpaper, 10 pt, conference]{ieeeconf}  

\IEEEoverridecommandlockouts                              

\usepackage{amsmath,amssymb,enumerate,color,soul}
\usepackage{epstopdf}
\usepackage{times} 
\usepackage[dvips]{graphicx}
\usepackage{psfrag}
\usepackage{multirow} 
\usepackage{cite} 
\usepackage{float}
\usepackage{colortbl}
\usepackage{stfloats}

\newtheorem{thm}{Theorem}

\newtheorem{lemma}{Lemma}

\newtheorem{clm}{Claim}

\newcommand{\arctanh}{\operatorname{arctanh}}
\newcommand{\dom}{\ensuremath{\text{dom}\,}}
\newcommand{\avg}{\ensuremath{\text{avg}\,}}
\newcommand{\diag}{\ensuremath{\text{diag}\,}}
\newcommand{\R}[2]{\ensuremath{\mathbb{R}^{#1}_{#2}}}
\newcommand{\Zp}{\ensuremath{\mathbb{Z}_{\geq 0}}}
\newcommand{\Zg}{\ensuremath{\mathbb{Z}_{> 0}}}

\newcommand{\KL}{\ensuremath{\mathcal{KL}}}

\graphicspath{{Figures/}}
\newcounter{MYtempeqncnt}

\title{\LARGE \bf
Co-design of output feedback laws and event-triggering conditions for linear systems
}

\author{Mahmoud Abdelrahim, Romain Postoyan, Jamal Daafouz and Dragan Ne\v{s}i\'c
\thanks{M. Abdelrahim, R. Postoyan and J. Daafouz are with the Universit\'{e} de Lorraine, CRAN, UMR 7039 and the CNRS, CRAN, UMR 7039, France {\tt\small \{othmanab1,romain.postoyan,jamal.daafouz\}@} {\tt\small univ-lorraine.fr}. Their work is partially supported by the ANR under the grant COMPACS (ANR-13-BS03-0004-02).}
\thanks{D. Ne\v{s}i\'c is with the Department of Electrical and Electronic Engineering, the University of Melbourne, Parkville, VIC 3010, Australia {\tt\small dnesic@unimelb.edu.au}. His work is supported by the Australian Research Council under the Discovery Projects and Future Fellowship schemes.}
}

\begin{document}

\maketitle
\thispagestyle{empty}
\pagestyle{empty}

\begin{abstract}
We present a procedure to simultaneously design the output feedback law and the event-triggering condition to stabilize linear systems. The closed-loop system is shown to satisfy a global asymptotic stability property and the existence of a strictly positive minimum amount of time between two transmissions is guaranteed. The event-triggered controller is obtained by solving linear matrix inequalities (LMIs). We then exploit the flexibility of the method to maximize the guaranteed minimum amount of time between two transmissions. Finally, we provide a (heuristic) method to reduce the amount of transmissions, which is supported by numerical simulations.
\end{abstract}

\section{Introduction}
Networked control systems (NCS) and embedded systems are becoming essential in a wide range of control applications. A crucial challenge for these systems is the efficient use of their limited resources in terms of communication and/or computation. In this context, event-triggered control has been proposed as an alternative to the conventional periodic sampling paradigm. The idea is to close the loop and update the control input whenever a state-dependent criterion is verified, which is designed based on the stability/performance requirements, see \textit{e.g.} \cite{Arzen1999simple, Astrom1999comparison, Tabuada2007event, Wang2011Event, Romain2011unifying, Heemels2012Introduction}. In this way, it is possible to significantly reduce the usage of the communication resources by the control task compared to periodic sampling. In this paper, we will focus on the scenario where we want to reduce the amount of control updates, which is relevant in the context of networked control systems for instance as this leads to a reduced usage of the network, which can thus be used by other tasks.

The vast majority of existing event-triggered controllers are designed by emulation, see \cite{Heemels2012Introduction, Romain2011unifying, Wang2011Event} and the references therein. In other words, a stabilizing feedback law is first constructed in the absence of network and then the triggering condition is synthesized to preserve stability. The potential disadvantage of this technique is that it is difficult to obtain an \textit{optimal} design since we are restricted by the initial choice of the feedback law. To overcome this issue, three directions of research are proposed in the literature: joint design of control inputs and self-triggering conditions, \textit{e.g.} \cite{Donkers2011on,Gommans2014self}, optimal event-triggered control, \textit{e.g.} \cite{Antunes2012dynamic,Molin2010optimal}, and co-design of feedback laws and event-triggering conditions, \textit{e.g.} \cite{Hu2012event,Shanbin2011codesign,Peng2013event}. In this paper, we are interested in the last approach.

All of the aforementioned results focus on state feedback event-triggered controllers where the full state vector is assumed to be available for measurement. However, from a practical point of view, this is not realistic for many control applications where only a part of the plant is measured. It is important to highlight that the design of event-triggered controllers based on the output measurements is particularly challenging, even by emulation, see \cite{Kofman2006level,Donkers2012output,Peng2013output,Tallapragada2012event-CDC,Forni2014event,Yu2012event,Abdelrahim2014stabilization}. This is due to the fact that it is usually difficult to ensure the existence of a uniform strictly positive lower bound on the inter-transmission times contrary to the case where the full state is measured (see\cite{Donkers2012output}).

The purpose of this paper is to develop a joint design procedure of the output feedback law and the event-triggering condition. To the best of the authors' knowledge, this problem has been only addressed in \cite{Zhang2013event}, \cite{Meng2014event}. The proposed co-design methods in \cite{Zhang2013event}, \cite{Meng2014event} are concerned with periodic event-triggered controllers (PETC) in which the output measurements are sampled periodically and then it is the task of the triggering condition to decide whether the control input needs to be updated. However, an open question regarding these techniques is how to calculate the appropriate sampling period of the triggering mechanism. This is a key aspect in the construction of PETC since the sampling of the triggering mechanism may deteriorate the closed-loop performance or may require a higher network bandwidth than the available one, see \cite{Romain2013periodic}.

Unlike \cite{Zhang2013event}, \cite{Meng2014event}, we provide a co-design algorithm where the triggering condition is continuously evaluated. The scheme is based on our previous work in \cite{Abdelrahim2014stabilization} where we have synthesized stabilizing output feedback event-triggered controllers by emulation. The proposed triggering mechanism in \cite{Abdelrahim2014stabilization} guarantees a global asymptotic stability property for the closed-loop and enforces a minimum amount of time $T$ between two transmission instants by combining the event-triggering condition of \cite{Tabuada2007event} (adapted to output feedbacks) and the time-triggered results in \cite{Nesic2009explicit}. The constant time $T$ corresponds to the \textit{maximum allowable sampling period} (MASP) given by \cite{Nesic2009explicit}. Contrary to \cite{Abdelrahim2014stabilization}, where the output feedback laws were assumed to be known a priori, in this paper, we simultaneously design the feedback controllers and the transmission rules for linear time-invariant (LTI) systems. The event-triggered controller is then obtained by solving LMIs. It is important to note that the results in \cite{Abdelrahim2014stabilization} do not allow for co-design because the resulted LMI condition is nonlinear in this case. Furthermore, the encountered nonlinearity cannot be directly handled by congruence transformations like in standard output feedback design problems, which induces non-trivial technical difficulties. We thus needed to introduce an additional LMI constraint to linearize the LMI condition in \cite{Abdelrahim2014stabilization} using the tools of \cite{Scherer1997multiobjective}. We then take advantage of the flexibility of co-design to enhance the efficiency of the event-triggered controllers in two senses. We first maximize the minimum inter-transmission time which is essential in practice. Indeed, while the existence of dwell-time is typically ensured in emulation results, its value may be very small and may thus violate the hardware constraints. It is therefore important to propose designs which are able to ensure larger minimum times between two transmissions. We then propose a heuristic to reduce the amount of transmissions, whose efficiency is confirmed by simulations.

The rest of the paper is organised as follows. Preliminaries are given in Section \ref{sec: preliminaries}. The problem is formally stated in Section \ref{sec: problem-statement}. In Section \ref{sec: main-results}, we give the main results and in Section \ref{sec: optimaztion-problems} we explain how these results can be used to enlarge the guaranteed minimum inter-transmission time and to reduce the amount of transmissions. An illustrative example is proposed in Section \ref{sec: illutrative-example}. Conclusions are provided in Section \ref{sec: conclusions}. The proofs are given in the Appendix.

\section{Preliminaries} \label{sec: preliminaries}
Let $\R{}{} := (-\infty,\infty)$, $\R{}{\geq 0} := [0,\infty)$, $\Zp := \{ 0, 1, 2, . . \}$ and $\Zg := \{ 1, 2, . . \}$. We denote the minimum and maximum eigenvalues of the symmetric matrix $A$ as $\lambda_{\min}(A)$ and $\lambda_{\max}(A)$, respectively. We write $A^{T}$ and $A^{-T}$ to respectively denote the transpose and the inverse of transpose of $A$ and $\diag(A_{1},\cdots,A_{N})$ is the block-diagonal matrix with the entries $A_{1},\cdots,A_{N}$ on the diagonal. The symbol $\star$ stands for symmetric blocks. We use $\mathbb{I}_{n}$ to denote the identity matrix of dimension $n$. The shorthand $\Sigma(Q)$ stands for $Q + Q^{T}$ for any square matrix $Q$. The Euclidean norm is denoted as $|.|$. We use $(x,y)$ to represent the vector $[x^{T}, y^{T}]^{T}$ for $x \in \R{n}{}$ and $y \in \R{m}{}$.

In this paper, we consider hybrid systems of the following form using the formalism of \cite{Teel}
\begin{equation}\label{eq: Prelim-hybrid-model}
  \dot{x} = F(x) \hspace{10pt} x\in C, \hspace{20pt} x^{+} = G(x) \hspace{10pt} x\in D,
\end{equation}
where $x\in \R{n}{}$ is the state, $F$ is the flow map, $C$ is the flow set, $G$ is the jump map and $D$ is the jump set. The vector fields $F$ and $G$ are assumed to be continuous and the sets $C$ and $D$ are closed. The solutions to system (\ref{eq: Prelim-hybrid-model}) are defined on so-called hybrid time domains. A set $E \subset \R{}{\geq0}\times \Zp$ is called a \textit{compact hybrid time domain} if $E =  \displaystyle \underset{j\in\{0,...,J-1\}}{\cup}([t_{j}, t_{j+1}], j)$ for some finite sequence of times $0=t_{0}\leq t_{1} \leq ... \leq t_{J}$ and it is a \textit{hybrid time domain} if for all $(T,J)\in E, E \cap ([0,T]\times \{0,1,...,J\})$ is a compact hybrid time domain. A function $\phi:E\rightarrow\R{n}{}$ is a hybrid arc if $E$ is a hybrid time domain and if for each $j\in\Zp, t \mapsto \phi(t,j)$ is locally absolutely continuous on $I^{j} := \{ t: (t,j)\in E \}$. A hybrid arc $\phi$ is a solution to system (\ref{eq: Prelim-hybrid-model}) if: (i) $\phi(0,0)\in C\cup D$; (ii) for any $j\in \Zp$, $\phi(t,j) \in C$ and $\dot{\phi}(t,j) = F(\phi(t,j))$ for almost all $t \in I^{j}$; (iii) for every $(t,j)\in \dom \phi$ such that $(t,j+1)\in \dom \phi$, $\phi(t,j) \in D$ and $\phi(t,j+1) = G(\phi(t,j))$. A solution $\phi$ to system (\ref{eq: Prelim-hybrid-model}) is \textit{maximal} if it cannot be extended, and it is \textit{complete} if its domain, $\dom \phi$, is unbounded.

\section{Problem statement} \label{sec: problem-statement}
\begin{figure}[h!]
\centering \scriptsize
\psfrag{Plant}[][][1]{Plant}
\psfrag{ETC}[][][1]{Event-triggering}
\psfrag{mechanism}[][][1]{mechanism}
\psfrag{Controller}[][][1]{Controller}
\psfrag{yt}[][][1]{$y(t)$}
\psfrag{yk}[][][1]{$y(t_{i})$}
\psfrag{ut}[][][1]{$u(t)$}
\psfrag{uk}[][][1]{$u(t_{i})$}
\includegraphics[scale=0.3]{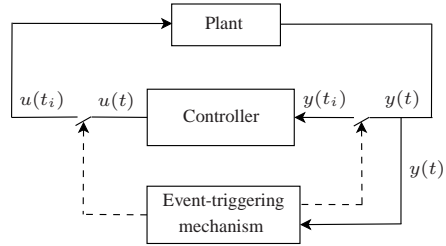}
\caption{Event-triggered control schematic}\label{fig:output-controller}
\end{figure}

Consider the linear time-invariant system
\begin{equation}\label{plant-linear-ex-out}
\dot{x}_{p} = A_{p}x_{p} + B_{p}u, \hspace{30pt} y = C_{p}x_{p},
\end{equation}
where $x_{p}\in \R{n_{p}}{}$, $u\in \R{n_{u}}{}$, $y\in \R{n_{y}}{}$ and $A_{p}, B_{p}, C_{p}$ are matrices of appropriate dimensions. We will design dynamic output feedback laws of the form
\begin{equation}\label{controller-linear-ex-out}
\dot{x}_{c} = A_{c}x_{c} + B_{c}y, \hspace{30pt} u = C_{c}x_{c},
\end{equation}
where $x_{c} \in \R{n_{c}}{}$ and $A_{c}, B_{c}, C_{c}$ are matrices of appropriate dimensions. We focus on the case where the controller has the same dimension as the plant, \textit{i.e.} $n_{c} = n_{p}$. We emphasize that the $x_{c}$-system is not necessarily an observer. We consider the scenario where controller (\ref{controller-linear-ex-out}) communicates with the plant via a digital channel. Hence, the plant output and the control input are sent only at transmission instants $t_{i}, i\in \Zp$. In this paper, we are interested in an event-triggered implementation in the sense that the sequence of transmission instants is determined by a criterion based on the output measurements, like in \cite{Donkers2012output,Tallapragada2012event-CDC}, see Figure \ref{fig:output-controller}.

\begin{figure*}[!t]
\normalsize
\setcounter{MYtempeqncnt}{\value{equation}}

\setcounter{equation}{8}
\begin{equation}\label{eq: prop-ineq1}
\begin{array}{l}
\left( \begin{array}{ccccccccc}
\Sigma(\boldsymbol YA_{p} + \boldsymbol ZC_{p}) &\star &\star & \star &\star &\star &\star \\
A_{p} + \boldsymbol M^{T} & \Sigma(A_{p}\boldsymbol X + B_{p}\boldsymbol N) &\star & \star &\star &\star &\star \\
\boldsymbol Z^{T} & 0 & -\boldsymbol\mu\mathbb{I}_{n_{y}} & \star &\star &\star &\star \\
B_{p}^{T}\boldsymbol Y & B_{p}^{T} & 0  & -\boldsymbol\mu\mathbb{I}_{n_{u}} &\star &\star &\star  \\
\boldsymbol YA_{p} + \boldsymbol ZC_{p} & \boldsymbol M & 0 & 0 & -\boldsymbol Y &\star &\star \\
A_{p} & A_{p}\boldsymbol X + B_{p}\boldsymbol N & 0 & 0 & -\mathbb{I}_{n_{p}} & -\boldsymbol X &\star  \\
C_{p} & C_{p}\boldsymbol X & 0 &  0 & 0 & 0 & -\boldsymbol \varepsilon\mathbb{I}_{n_{y}}
\end{array} \right)  < 0
\end{array}
\end{equation}
\setcounter{equation}{\value{MYtempeqncnt}}
\hrulefill
\vspace*{4pt}

\end{figure*}

\noindent At each transmission instant, the plant output is sent to the controller which computes a new control input that is instantaneously transmitted to the plant. We assume that this process is performed in a synchronous manner and we ignore the computation times and the possible transmission delays. In that way, like in \cite{Daci2007quadratic}, we obtain
\begin{equation}\label{eq: system-with-network}
  \begin{array}{rcll}
    \dot{x}_{p} &=& A_{p}x_{p} + B_{p}\hat{u} &\hspace{30pt} t\in [t_{i}, t_{i+1}]\\
    \dot{x}_{c} &=& A_{c}x_{c} + B_{c}\hat{y} &\hspace{30pt} t\in [t_{i}, t_{i+1}]\\
    u &=& C_{c}x_{c} \\
    \dot{\hat{y}} &=& 0 &\hspace{30pt} t\in [t_{i}, t_{i+1}]\\
    \dot{\hat{u}} &=& 0 &\hspace{30pt} t\in [t_{i}, t_{i+1}] \\
    \hat{y}(t_{i}^{+}) &=& y(t_{i}) \\
    \hat{u}(t_{i}^{+}) &=& u(t_{i}),
  \end{array}
\end{equation}
\noindent where $\hat{y}$ and $\hat{u}$ respectively denote the last transmitted values of the plant output and the control input. We assume that zero-order-hold devices are used to generate the sampled values $\hat{y}$ and $\hat{u}$, which leads to  $\dot{\hat{y}} = 0$ and $\dot{\hat{u}} = 0$. We introduce the network-induced error $e := (e_{y}, e_{u})\in \R{n_{e}}{}$, where $e_{y} := \hat{y} - y$, $e_{u} := \hat{u} - u$ and $n_{e} = n_{y} + n_{u}$
which are reset to $0$ at each transmission instant. We model the event-triggered control system using the hybrid formalism of \cite{Teel} as in \emph{e.g.} \cite{Donkers2012output}, \cite{Forni2014event}, \cite{Romain2011unifying}, for which a jump corresponds to a transmission. In that way, the system can be modeled as
\begin{equation} \label{eq: hybrid-model}
\arraycolsep=2.6pt\def\arraystretch{1.2}
\begin{array}{rcll}
\left(\begin{array}{c} \dot{x}\\ \dot{e}\\ \dot{\tau}\end{array} \right) &=& \left(\begin{array}{c} \mathcal{A}_{1}x + \mathcal{B}_{1}e\\ \mathcal{A}_{2}x +  \mathcal{B}_{2}e\\ 1\end{array} \right) & (x,e,\tau)\in C \\[15pt]
\left(\begin{array}{c} x^{+}\\ e^{+}\\ \tau^{+}\end{array} \right) &=& \left(\begin{array}{c} x\\ 0\\ 0\end{array} \right) & (x,e,\tau) \in D,
\end{array}
\end{equation}
where $x = (x_{p}, x_{c})\in \R{n_{x}}{}$ with $n_{x} = 2n_{p}$, $\tau \in \R{}{\geq 0}$ is a clock variable which describes the time elapsed since the last jump and
\begin{equation} \label{xdot-linear-ex-out}
\arraycolsep=0.8pt\def\arraystretch{1.2}
\begin{array}{rllrlll}
\mathcal{A}_{1} &=& \left(\begin{array}{cc} A_{p} & B_{p}C_{c}\\ B_{c}C_{p} & A_{c} \end{array}\right) &\hspace{6pt} \mathcal{B}_{1} &=&\left(\begin{array}{cc} 0 & B_{p}\\ B_{c} & 0 \end{array}\right) \\ [12pt]
\mathcal{A}_{2} &=& \left(\begin{array}{cc} - C_{p}A_{p} & - C_{p}B_{p}C_{c}\\ - C_{c}B_{c}C_{p} & - C_{c}A_{c} \end{array}\right) &\hspace{6pt} \mathcal{B}_{2} &=&\left(\begin{array}{cc} 0 & - C_{p}B_{p}\\ -C_{c}B_{c} & 0 \end{array}\right).
\end{array}
\end{equation}
The flow and jump sets of (\ref{eq: hybrid-model}) are defined according to the triggering condition we will design in the next section. As long as the triggering condition is not violated, the system flows on $C$ and a jump occurs when the state enters in $D$. When $(x,e,\tau) \in C \cap D$, the solution may flow only if flowing keeps $(x,e,\tau)$ in $C$, otherwise the system experiences a jump. The sets $C$ and $D$ will be closed (which ensure that system (\ref{eq: hybrid-model}) is well-posed, see Chapter 6 in \cite{Teel}).

The main objective of this paper is to simultaneously design the dynamic controller (\ref{controller-linear-ex-out}) and the flow and the jump sets of system (\ref{eq: hybrid-model}), \textit{i.e.} the triggering condition, to ensure a global asymptotic stability property for system (\ref{eq: hybrid-model}).

\section{Main results}\label{sec: main-results}
We use the same triggering condition as in \cite{Abdelrahim2014stabilization}, \emph{i.e.}
\begin{equation}\label{flow-jump-sets-out}
\arraycolsep=1.4pt\def\arraystretch{2.2}
\begin{array}{lcll}
C &=& \Big\{(x,e,\tau): &\gamma^{2} |e|^{2} \leq \varepsilon_{1}|y|^{2} \text{ or } \tau \in [0,T]\Big\} \\
D &=& \Big\{(x,e,\tau): &\Big(\gamma^{2} |e|^{2} = \varepsilon_{1}|y|^{2} \text{ and } \tau \geq T\Big) \text{ or } \\
 & & &\Big(\gamma^{2} |e|^{2} \geq \varepsilon_{1}|y|^{2} \text{ and } \tau = T\Big) \Big\},
\end{array}
\end{equation}
where $\gamma\geq 0$, $\varepsilon_{1}>0$ are design parameters and $T$ is a constant which enforces a uniform dwell-time between any two jumps. This constant $T$ is designed such that $T < \mathcal{T}(\gamma, L)$, where $\mathcal{T}(\gamma, L)$ corresponds to the maximum allowable sampling period given in \cite{Nesic2009explicit}, which is given by
\begin{equation} \label{T-phi}
\mathcal{T}(\gamma, L) := \left\{
    \begin{array}{ll}
    \frac{1}{Lr}\arctan(r)& \hspace{10pt} \gamma > L \\[4pt]
    \frac{1}{L}& \hspace{10pt} \gamma = L \\[4pt]
    \frac{1}{Lr}\arctanh(r)& \hspace{10pt} \gamma < L
    \end{array}
    \right.
\end{equation}
where $r := \sqrt{\left|(\frac{\gamma}{L})^{2} - 1\right|}$ and $L := |\mathcal{B}_{2}|$.

The following theorem provides LMI-based conditions to simultaneously design the output feedback law (\ref{controller-linear-ex-out}) and the parameters of the flow and jump sets (\ref{flow-jump-sets-out}) such that a global asymptotic stability property holds for system (\ref{eq: hybrid-model}),  (\ref{flow-jump-sets-out}). We use boldface symbols to emphasize the LMIs decision variables.

\vspace{0.2cm}
\begin{thm}\label{thm: linear-systems-co-design}
Consider system (\ref{eq: hybrid-model}) with the flow and jump sets (\ref{flow-jump-sets-out}). Suppose that there exist symmetric positive definite real matrices $\boldsymbol X, \boldsymbol Y\in\R{n_{p}\times n_{p}}{}$, real matrices $\boldsymbol M\in\R{n_{p}\times n_{p}}{}, \boldsymbol Z\in\R{n_{p}\times n_{y}}{}, \boldsymbol N\in\R{n_{u}\times n_{p}}{}$ and $\boldsymbol\varepsilon, \boldsymbol\mu > 0$ such that (\ref{eq: prop-ineq1}) is verified and the following holds

\setcounter{equation}{9}
\begin{equation}\label{eq: prop-ineq2}
\left( \begin{array}{ccccccccc}
-\mathbb{I}_{n_{y}} &\star &\star &\star \\
0 & -\mathbb{I}_{n_{u}} &\star &\star \\
-C_{p}^{T} & 0 & -\boldsymbol Y &\star \\
-\boldsymbol XC_{p}^{T} & -\boldsymbol N^{T} & -\mathbb{I}_{n_{p}} & -\boldsymbol X
\end{array} \right) < 0.
\end{equation}
Take $\gamma = \sqrt{\boldsymbol\mu}$, $\varepsilon_{1}  = \boldsymbol\varepsilon^{-1}$ and
\begin{equation}\label{eq: prop-controller}
\arraycolsep=1.4pt\def\arraystretch{1.5}
\begin{array}{lllllll}
A_{c} &=& V^{-1}(\boldsymbol M - \boldsymbol YA_{p}\boldsymbol X - \boldsymbol YB_{p}\boldsymbol N - \boldsymbol ZC_{p}\boldsymbol X)U^{-T}\\
B_{c} &=& V^{-1}\boldsymbol Z, \hspace{15pt} C_{c} = \boldsymbol NU^{-T},
\end{array}
\end{equation}
where $U, V \in \R{n_{p}\times n_{p}}{}$ are any square and invertible matrices such that\footnote{In view of the Schur complement of LMI (\ref{eq: prop-ineq2}), we deduce that $\left( \begin{smallmatrix} \boldsymbol Y & \mathbb{I}_{n_{p}} \\ \mathbb{I}_{n_{p}} & \boldsymbol X\end{smallmatrix}\right)>0$
which implies that $\boldsymbol X - \boldsymbol Y^{-1}>0$ and thus, $\mathbb{I}_{n_{p}} - \boldsymbol X\boldsymbol Y$ is nonsingular. Hence, the existence of nonsingular matrices $U, V$ is always ensured.} $UV^{T} = \mathbb{I}_{n_{p}} - \boldsymbol X\boldsymbol Y$. Then, there exists $\chi \in \KL$ such that any solution $\phi = (\phi_{x},\phi_{e},\phi_{\tau})$ satisfies
\begin{equation} \label{eq-thm-clock-out}
|\phi_{x}(t,j)| \leq \chi(|(\phi_{x}(0,0), \phi_{e}(0,0))|, t + j) \hspace{10pt} \forall(t,j) \in \dom \phi
\end{equation}
and, if $\phi$ is maximal, it is also complete.
\hfill $\Box$
\end{thm}
\vspace{0.2cm}
We note that LMIs (\ref{eq: prop-ineq1}), (\ref{eq: prop-ineq2}) are computationally tractable and can be solved using the SEDUMI solver \cite{Sturm1999sedumi} with the YALMIP interface \cite{Lofberg2004Yalmip}. Hence, by solving (\ref{eq: prop-ineq1}) and (\ref{eq: prop-ineq2}), we obtain the feedback law, see (\ref{eq: prop-controller}), and the triggering condition parameters $\gamma$ and $\varepsilon_{1}$.

The proof of Theorem \ref{thm: linear-systems-co-design} consists in showing that the following holds
\begin{equation}\label{eq: prop-LMI}
\arraycolsep=1.4pt\def\arraystretch{1.2}
\left( \begin{array}{cccc} \mathcal{A}_{1}^{T}\boldsymbol P\!+\! \boldsymbol P\mathcal{A}_{1}\! + \! \mathcal{A}_{2}^{T}\mathcal{A}_{2}\! + \!\boldsymbol\varepsilon_{1}\overline{C}_{p}^{T}\overline{C}_{p} & \star \\ \mathcal{B}_{1}^{T}\boldsymbol P & -\boldsymbol\mu\mathbb{I}_{n_{e}} \end{array} \right) < 0,
\end{equation}
where $\boldsymbol P$ is the Lyapunov matrix and $\overline{C}_{p} = [C_{p} \hspace{12pt} 0]$. The LMI (\ref{eq: prop-LMI}) corresponds to the condition in Proposition 1 in \cite{Abdelrahim2014stabilization} in the context of emulation, \textit{i.e.} when the controller is given. It is important to note that the derivation of LMIs for co-design from (\ref{eq: prop-LMI}) is not trivial, because of the nonlinear term $\mathcal{A}_{2}^{T}\mathcal{A}_{2}$ which depends on the controller matrices. This term never appeared in the classical output feedback design problems and it is the reason why the LMI (\ref{eq: prop-ineq1}) differs from the classical one and why the additional convex constraint (\ref{eq: prop-ineq2}) is needed in Theorem \ref{thm: linear-systems-co-design}.


\section{Optimization problems}\label{sec: optimaztion-problems}
The flexibility of the co-design procedure proposed in Section \ref{sec: main-results} can be exploited in many ways. In this section, we explain how to use the LMI conditions (\ref{eq: prop-ineq1}) and (\ref{eq: prop-ineq2}) to enlarge the guaranteed minimum amount of time between any two transmissions. We then propose a heuristic method to reduce the amount of transmissions. The efficiency of these methods is illustrated by simulations in Section \ref{sec: illutrative-example}.

\subsection{Enlarging the guaranteed minimum inter-transmission time}\label{sec: optim-T}
A key challenge in the design of output feedback event-triggered controllers is to ensure the existence of a uniform strictly positive lower bound on the inter-transmission times. Although the existence of that lower bound is guaranteed by different techniques in the literature, the available expressions are often subject to some conservatism. It is therefore unclear whether the event-triggered controller has a dwell-time which is compatible with the hardware limitations. We investigate in this section how to employ the LMIs conditions (\ref{eq: prop-ineq1}), (\ref{eq: prop-ineq2}) to maximize the guaranteed minimum inter-transmission time. We first state the following lemma to motivate our approach.
\vspace{0.2cm}
\begin{lemma}\label{lma-T-min}
Let $\mathcal{S}$ be the set of solutions to system (\ref{eq: hybrid-model}), (\ref{flow-jump-sets-out}).
\begin{equation}\label{eq:lma-T-min}
\begin{array}{rrr}
  T =\underset{\phi\in\mathcal{S}}\inf\{t'-t: \exists j \in \Zg, \,\, (t,j), (t,j+1), (t',j+1), \\(t',j+2)\in \dom\phi\}.
\end{array}
\end{equation}
\hfill $\Box$
\end{lemma}
Lemma \ref{lma-T-min} implies that the lower bound $T$ on the inter-transmission times guaranteed by (\ref{flow-jump-sets-out}) corresponds to the actual minimum inter-transmission time as defined by the right-hand side of (\ref{eq:lma-T-min}). Hence, by maximizing $T$, we enlarge the minimum inter-transmission time.

To maximize $T$, we will maximize $\mathcal{T}(\gamma, L)$ in (\ref{T-phi}). We see that $\mathcal{T}$ increases as $\gamma$ and $L$ decrease. Hence, our objective is to minimize $\gamma$ and $L$. Since $\gamma$ corresponds to $\sqrt{\mu}$ and $\mu$ enters linearly in the LMI (\ref{eq: prop-ineq1}), we can directly minimize $\gamma$ under the LMIs constraints (\ref{eq: prop-ineq1}), (\ref{eq: prop-ineq2}). The minimization of $L$, on the other hand, requires more attention. We recall that $L = |\mathcal{B}_{2}| = \sqrt{\lambda_{\max}(\mathcal{B}_{2}^{T}\mathcal{B}_{2})}$, where
\begin{equation}\label{eq: B2TB2}
\mathcal{B}_{2}^{T}\mathcal{B}_{2} = \left(\begin{array}{cc}
B_{c}^{T}C_{c}^{T}C_{c}B_{c} & 0\\0 & B_{p}^{T}C_{p}^{T}C_{p}B_{p}
\end{array}
\right)
\end{equation}
hence,
\begin{equation} \label{eq: L}
  L = \max\left(\sqrt{\lambda_{\max}(B_{c}^{T}C_{c}^{T}C_{c}B_{c})}, \sqrt{\lambda_{\max}(B_{p}^{T}C_{p}^{T}C_{p}B_{p})}\right).
\end{equation}
Therefore, $L$ can be minimized up to $\sqrt{\lambda_{\max}(B_{p}^{T}C_{p}^{T}C_{p}B_{p})}$ which is fixed as it only depends on the plant matrices. In view of (\ref{eq: prop-controller}), we have that
\begin{equation}
  B_{c}^{T}C_{c}^{T}C_{c}B_{c} = \boldsymbol Z^{T}V^{-T}U^{-1}\boldsymbol N^{T}\boldsymbol NU^{-T}V^{-1}\boldsymbol Z.
\end{equation}
Thus, $L$ depends nonlinearly on the LMI variables $\boldsymbol N$ and $\boldsymbol Z$ and it can a priori not be directly minimized. To overcome this issue, we impose the following upper bound
\begin{equation}\label{alpha}
  B_{c}^{T}C_{c}^{T}C_{c}B_{c} < \alpha\beta\mathbb{I}_{n_{y}}
\end{equation}
for some $\alpha, \beta>0$. As a result, to minimize $\alpha$ and $\beta$ may help to minimize $L$ as we will show on an example in Section \ref{sec: illutrative-example}. We translate inequality (\ref{alpha}) into a LMI constraint and we state the following claim.
\begin{clm} \label{clm: LMI3}
Assume that LMIs (\ref{eq: prop-ineq1}), (\ref{eq: prop-ineq2}) are verified. Then, there exist $\boldsymbol\alpha, \boldsymbol\beta>0$ such that
\begin{equation}\label{eq: alpha-LMI}
\left( \begin{array}{ccccccccc}
\boldsymbol \alpha\mathbb{I}_{n_{y}} &\star &\star &\star \\
0 & \boldsymbol \beta\mathbb{I}_{n_{u}} &\star &\star \\
0 & \boldsymbol N^{T} & \boldsymbol X &\star \\
\boldsymbol Z & 0 & \mathbb{I}_{n_{p}} & \boldsymbol Y
\end{array} \right) > 0
\end{equation}
which implies that inequality (\ref{alpha}) holds.
\hfill $\Box$
\end{clm}
\vspace{0.2cm}

We note that (\ref{eq: alpha-LMI}) does not introduce additional constraints on system (\ref{eq: hybrid-model}) compared to LMIs (\ref{eq: prop-ineq1}), (\ref{eq: prop-ineq2}). This comes from the fact that there always exist $\alpha, \beta>0$ (eventually large) such that (\ref{eq: alpha-LMI}) holds, in view of Schur complement of (\ref{eq: alpha-LMI}).

In conclusion, we formulate the problem as a multiobjective optimization problem as we want to minimize $\mu, \alpha, \beta$ under the constraint (\ref{eq: prop-ineq1}), (\ref{eq: prop-ineq2}) and (\ref{eq: alpha-LMI}). Several approaches have been proposed in the literature to handle that kind of problems, see \textit{e.g.} \cite{Ehrgott}. We choose the weighted sum strategy among others and we formulate the LMI optimization problem as follows
\begin{equation} \label{eq: codesign-optim-T}
\underset{\textstyle \text{subject to (\ref{eq: prop-ineq1}), (\ref{eq: prop-ineq2}), (\ref{eq: alpha-LMI})}}{\min{\lambda_{1}\mu + \lambda_{2}\alpha} + \lambda_{3}\beta}
\end{equation}
for some weights $\lambda_{1}, \lambda_{2}, \lambda_{3} \geq 0$.

\subsection{Reducing the amount of transmissions}\label{sec: optim-trans}
We present a heuristic way to reduce the amount of transmissions generated by the triggering mechanism. This goal can be achieved by optimizing the parameters of the event-triggered rule such that the triggering condition is violated after the longest possible time since the last transmission. In view of (\ref{flow-jump-sets-out}) and Theorem \ref{thm: linear-systems-co-design}, since $\gamma = \sqrt{\mu}, \varepsilon_{1} = \varepsilon^{-1}$, the event-triggering condition is given by
\begin{equation}\label{trig-cond}
  \mu|e|^{2} \leq \varepsilon^{-1}|y|^{2} \text{ or } \tau \in [0,T].
\end{equation}
As a consequence, in order to reduce the number of instants at which the rule (\ref{trig-cond}) is not satisfied, we need to minimize the parameters $\mu$ and $\varepsilon$. More precisely, we need to minimize the product $\varepsilon\mu$. Since the product $\varepsilon\mu$ is nonlinear, we simply minimize the weighted sum of the two parameters to maintain the convexity property. Moreover, we need to take into account the evolution of the $e$-variable. Indeed, it is not because $\varepsilon\mu$ is minimized that less transmissions will occur because the variable $e$ may more rapidly reach the threshold in (\ref{trig-cond}) in this case. To address this point, we notice that, in view of Assumption 1 and Proposition 1 in \cite{Abdelrahim2014stabilization}, the variable $e$ satisfies, for all $x \in \R{n_{x}}{}$ and almost all $e \in \R{n_{e}}{}$
\begin{align}\label{W-dot-out}
  \langle\nabla |e|, \mathcal{A}_{2}x + \mathcal{B}_{2}e\rangle \leq L|e| + |\mathcal{A}_{2}x|.
\end{align}
Thus, minimizing $L$ may lead to the reduction of the rate of growth of the norm of the error.

To summarize, the optimization problem below may be used to reduce the amount of transmissions
\begin{equation}\label{eq: codesign-optim-T-eps2}
\underset{\textstyle \text{subject to (\ref{eq: prop-ineq1}), (\ref{eq: prop-ineq2}), (\ref{eq: alpha-LMI})}}{\min \lambda_{1}\mu + \lambda_{2}\alpha + \lambda_{3}\beta + \lambda_{4}\varepsilon}
\end{equation}
for some weights $\lambda_{1}, \lambda_{2}, \lambda_{3}, \lambda_{4} \geq 0$.

\section{Illustrative example}\label{sec: illutrative-example}
In this section, we demonstrate the potential of the proposed optimization problems on Example 2 in \cite{Donkers2012output}. Consider the LTI plant model
\begin{equation}\label{ex-Dokers}
\begin{array}{rll}
  \dot{x}_{p} &=& \begin{bmatrix} 0 & 1\\ -2 & 3 \end{bmatrix}x_{p} + \begin{bmatrix} 0 \\ 1 \end{bmatrix} u \\[10pt]
  y &=& \begin{bmatrix} -1 & 4 \end{bmatrix} x_{p}.
\end{array}
\end{equation}

\begin{figure}[h!]
\centering
\includegraphics[scale=0.53]{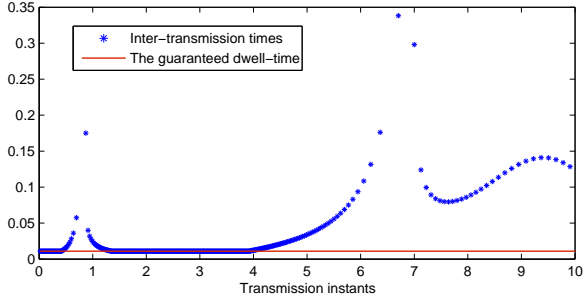}
\caption{Inter-transmission times for $(x(0),e(0),\tau(0)) = (10,-10,0,0,0,0,0)$.}\label{fig:Tau_Donkers}
\end{figure}

First, we solve the optimization problem (\ref{eq: codesign-optim-T}) to seek for the largest possible lower bound on the inter-transmission times. We set $\lambda_{1}=\lambda_{2}=\lambda_{3} = 1$ and we obtain
\begin{equation}
\begin{array}{lllllllllllllllll}
  T &=& 0.0114, & \mu &=& 18433, & \varepsilon &=& 2.7709\times10^{6} \\
  L &=& 4.0586, & \alpha &=& 4681.5, & \beta &=& 4.6599
\end{array}
\end{equation}
and
\begin{equation}\label{ex-Dokers-optim1}
\begin{array}{rlllll}
  A_{c} &=& \begin{bmatrix} 1.0919 & -1.1422\\ 4.9734 & -6.1425 \end{bmatrix}, & B_{c} = \begin{bmatrix} 16.7501 \\ 64.6472 \end{bmatrix},\\[10pt]
  C_{c} &=& \begin{bmatrix} 0.1157 & -0.0928 \end{bmatrix}.
\end{array}
\end{equation}
We note that, in view of (\ref{eq: L}), (\ref{ex-Dokers}), (\ref{ex-Dokers-optim1}), $L = \max(4.0855, 4) = 4.0855$. Table \ref{tbl:tau-min-avg-comp} gives the minimum and the average inter-sampling times, respectively denoted as $\tau_{\min}$ and $\tau_{\avg}$, for 100 randomly distributed initial conditions such that $|(x(0,0),e(0,0))|\leq 25$ and $\tau(0,0)=0$. The constant $\tau_{\avg}$ serves as a measure of the amount of transmissions (the bigger $\tau_{\avg}$, the less transmissions). We observe from the corresponding entries in Table \ref{tbl:tau-min-avg-comp} that $\tau_{\min} = \tau_{\avg}$ which implies that generated transmission instants are periodic. This may be explained by the fact that the product $\varepsilon\mu = 5.1075\times10^{10}$ is very big and thus the output-dependent part in (\ref{trig-cond}) is `quickly' violated. To avoid that phenomenon, we optimize the parameters of the event-triggering condition such that the rule is violated after the longest possible time since the last transmission instant, as discussed in Section \ref{sec: optim-trans}. Thus, we minimize the weighted sum $\lambda_{1}\mu + \lambda_{2}\alpha + \lambda_{3}\beta + \lambda_{4}\varepsilon$ subject to (\ref{eq: prop-ineq1}), (\ref{eq: prop-ineq2}), (\ref{eq: alpha-LMI}). We take $\lambda_{1}=\lambda_{2}=\lambda_{3}=\lambda_{4}=1$ and we obtain
\begin{equation}\label{eq:exmple-trig-param-optim1}
\begin{array}{lllllllllllllllll}
  T &=& 0.0113, & \mu &=& 18455, & \varepsilon &=& 28.6475 \\
  L &=& 4.0624, &\alpha &=& 4687.7, & \beta &=& 4.6669
\end{array}
\end{equation}
and the dynamic controller matrices are
\begin{equation}\label{eq:exmple-trig-contr-optim1}
\begin{array}{rlllll}
  A_{c} &=& \begin{bmatrix} 1.0927 & -1.1423\\ 4.9809 & -6.1477 & \end{bmatrix}, & B_{c} = \begin{bmatrix} 16.7530 \\ 64.7121 \end{bmatrix},\\[10pt]
  C_{c} &=& \begin{bmatrix} 0.1158 & -0.0927 \end{bmatrix}.
\end{array}
\end{equation}
We note from the corresponding entries in Table \ref{tbl:tau-min-avg-comp} that the guaranteed dwell-time $T$ is slightly smaller than the previous one but the average inter-transmission time $\tau_{\avg}$ is larger than the previous value (in this case $\varepsilon\mu = 5.2869\times10^{5}$). Furthermore, we can play with the weight coefficients $\lambda_{1}, \lambda_{2}, \lambda_{3}, \lambda_{4}$ to further reduce transmissions. Since we know that $L$ cannot become less than 4 and that the value obtained above is already close to this lower bound, we will give $\varepsilon$ the most relative importance by increasing the weight $\lambda_{4}$ to further decrease the magnitude of $\varepsilon\mu$. We found that the minimum value of $\varepsilon\mu = 8049$ is obtained with $\lambda_{1} = 1, \lambda_{2} = 0, \lambda_{3} = 0, \lambda_{4} = 10^{4}$ which yield
\begin{equation}
\begin{array}{lllllllllllllllll}
  T &=& 0.0109, & \mu &=& 19856, & \varepsilon &=& 0.4054 \\
  L &=& 4.3801, &\alpha &=& 8757, & \beta &=& 4418.3
\end{array}
\end{equation}
and the dynamic controller matrices are
\begin{equation}
\begin{array}{rlllll}
  A_{c} &=& \begin{bmatrix} 1.1684 & -1.1627\\ 5.6744 & -6.6241 & \end{bmatrix}, & B_{c} = \begin{bmatrix} 16.9843 \\ 70.3309 \end{bmatrix},\\[10pt]
  C_{c} &=& \begin{bmatrix} 0.1182 & -0.0908 \end{bmatrix}.
\end{array}
\end{equation}
We note that $\tau_{\avg}$ is twice bigger than with the controller (\ref{eq:exmple-trig-param-optim1}), (\ref{eq:exmple-trig-contr-optim1}) in this case and the guaranteed minimum inter-transmission time $T$ is of the same order of magnitude compared to the previous values, as shown in Table \ref{tbl:tau-min-avg-comp}. It is noted in Table \ref{tbl:tau-min-avg-comp} that, for all cases, the guaranteed lower bound $T$ corresponds to the minimum inter-transmission time $\tau_{\min}$ generated by the triggering mechanism. We provide the plot of the inter-transmission times for one simulation in Figure \ref{fig:Tau_Donkers} to better see the impact of the constant $T$ on the triggering instants.


In comparison, the guaranteed lower bound on the inter-transmission times in \cite{Donkers2012output} is $6.5\times10^{-9}$ while the observed lower bound and the average inter-transmission time during the simulations respectively are $4.8055\times10^{-6}$ and $2.2905\times10^{-4}$, as shown in Table \ref{tbl:tau-min-avg-comp}. Moreover, the stability property achieved in \cite{Donkers2012output} is a practical stability property, while we ensure a global asymptotic stability property. These observations justify the potential of the proposed co-design technique to reduce transmissions. In \cite{Meng2014event}, the guaranteed and the the simulated lower bounds on the inter-transmission times are found to be the sampling period $h=10^{-4}$, which is 100 times smaller than those we ensure.


\begin{table}[h!]
\begin{center}
\scalebox{0.8}{
\begin{tabular}{ c | c | c | c}
              & Guaranteed              & \multirow{2}{*}{$\tau_{\min}$}     & \multirow{2}{*}{$\tau_{\avg}$} \\
              &dwell-time & & \\\hline
  Donkers \& Heemels \cite{Donkers2012output} & \multirow{2}{*}{6.5$\times 10^{-9}$}  &     \multirow{2}{*}{$4.8055\times10^{-6}$}   & \multirow{2}{*}{$2.2905\times10^{-4}$}   \\
   $\sigma_{1} = \sigma_{2} = 10^{-3},\,\varepsilon_{1} = \varepsilon_{1} = 10^{-3}$ & & & \\ \hline
  Optimization problem (\ref{eq: codesign-optim-T})      & \multirow{2}{*}{0.0114} &  \multirow{2}{*}{0.0114} & \multirow{2}{*}{0.0114}  \\
  $\lambda_{1} = \lambda_{2} = \lambda_{3} = 1$ & & & \\ \hline
  Optimization problem (\ref{eq: codesign-optim-T-eps2}) & \multirow{2}{*}{0.0113} &  \multirow{2}{*}{0.0113} & \multirow{2}{*}{0.0116} \\
  $\lambda_{1} = 1, \lambda_{2} = 1, \lambda_{3} = 1, \lambda_{4} = 1$ & & & \\ \hline
  Optimization problem (\ref{eq: codesign-optim-T-eps2}) & \multirow{2}{*}{0.0109} &  \multirow{2}{*}{0.0109} & \multirow{2}{*}{0.0261} \\
  $\lambda_{1} = 1, \lambda_{2} = 0, \lambda_{3} = 0, \lambda_{4} = 10^{4}$ & & &
  \end{tabular}
  }
  \caption{Minimum and average inter-transmission times for 100 initial conditions in a ball of radius $25$ centered at the origin and for a simulation time of 20s.}
  \label{tbl:tau-min-avg-comp}
\end{center}
\end{table}

\section{Conclusion} \label{sec: conclusions}
A co-design procedure for output-feedback event-triggered controllers has been presented. LMIs conditions have been developed for that purpose. The proposed scheme guarantees a global asymptotic stability property for the closed-loop and enforces a strictly positive lower bound on the inter-transmission times. We have then used these LMIs to minimize transmissions between the plant and the controller in two different senses, while guaranteeing the closed-loop stability. In future work, we will further exploit these co-design results to take into account performance requirements.

\section*{Appendix}
\noindent \textbf{Proof of Theorem \ref{thm: linear-systems-co-design}.}
We define the following matrices
\begin{equation}
\arraycolsep=1.4pt\def\arraystretch{1.5}
\begin{array}{rllrll}
 S &=& \left(\begin{array}{cc} X & U \\ U^{T} & \hat{X} \end{array}\right), & S^{-1} &=& \left(\begin{array}{cc}  Y & V \\ V^{T} & \hat{Y} \end{array}\right) \\[15pt]
\Gamma &=& \left(\begin{array}{cc}  Y & \mathbb{I}_{n_{p}} \\ V^{T} & 0 \end{array}\right), & G &=& \left(\begin{array}{cc} -C_{p} & 0 \\ 0 & -C_{c} \end{array}\right),
 \end{array}
\end{equation}
where $\hat{X}, \hat{Y} \in \R{n_{p} \times n_{p}}{}$ are symmetric positive definite real matrices of appropriate dimension. Since $SS^{-1} = \mathbb{I}_{2n_{p}}$, it holds that $ X Y + UV^{T} = \mathbb{I}_{n_{p}}$, $ XV + U\hat{Y} = 0$, $U^{T} Y + \hat{X}V^{T} = 0$ and $U^{T}V + \hat{X}\hat{Y} = \mathbb{I}_{n_{p}}$. After some direct calculations, recall that $\overline{C}_{p} = [C_{p} \hspace{5pt} 0]$, we obtain
\begin{equation} \label{eq :prop-proof-calc1}
\arraycolsep=1.4pt\def\arraystretch{1.5}
\begin{array}{rll}
  S\Gamma &=& \left( \begin{array}{cc} \mathbb{I}_{n_{p}} &  X \\ 0 & U^{T} \end{array} \right), \hspace{15pt}
  \Gamma^{T}S\Gamma = \left( \begin{array}{cc}  Y & \mathbb{I}_{n_{p}} \\ \mathbb{I}_{n_{p}} &  X \end{array} \right) \\[15pt]
 \mathcal{B}_{1}^{T}\Gamma &=& \left( \begin{array}{cc} Z^{T} & 0 \\ B_{p}^{T} Y & B_{p}^{T} \end{array} \right), \hspace{11pt} GS\Gamma = \left(\begin{array}{cc} -C_{p} & -C_{p} X \\ 0 & - N \end{array}\right)\\[15pt]
\Gamma^{T}\mathcal{A}_{1}S\Gamma &=& \left(\begin{array}{cc}  YA_{p} + ZC_{p} &  M \\ A_{p} & A_{p} X + B_{p} N \end{array}\right)\\[15pt]
\overline{C}_{p}S\Gamma &=& (C_{p}\hspace{10pt}C_{p} X).
 \end{array}
\end{equation}
Consequently, inequalities (\ref{eq: prop-ineq1}), (\ref{eq: prop-ineq2}) can be written as
\begin{equation}\label{eq: prop-proof-ineq1}
\arraycolsep=1.4pt\def\arraystretch{1.5}
\begin{array}{c}
\left( \begin{array}{ccccccccc}
-\Gamma^{T}(S\mathcal{A}_{1}^{T} + \mathcal{A}_{1}S)\Gamma &\star &\star &\star \\
\mathcal{B}_{1}^{T}\Gamma & -\mu\mathbb{I}_{n_{e}} &\star &\star \\
\Gamma^{T}\mathcal{A}_{1}S\Gamma & 0 & -\Gamma^{T}S\Gamma &\star \\
\overline{C}_{p}S\Gamma & 0 & 0 & -\varepsilon\mathbb{I}_{n_{y}}
\end{array} \right)  <  0 \\[25pt]
\left( \begin{array}{ccccccccc}
-\mathbb{I}_{n_{e}} & GS\Gamma  \\
\Gamma^{T}SG^{T} & -\Gamma^{T}S\Gamma
\end{array} \right) < 0.
\end{array}
\end{equation}
By pre and post multiplying the first LMI respectively by $\diag(\mathbb{I}_{n_{x}}, \mathbb{I}_{n_{e}}, G\Gamma^{-T}, \mathbb{I}_{n_{y}})$ and its transpose and by using the Schur complement of the second LMI, we obtain
\begin{equation}\label{eq: prop-proof-ineq3}
\arraycolsep=1.4pt\def\arraystretch{1.5}
\begin{array}{c}
\left( \begin{array}{ccccccccc}
-\Gamma^{T}(S\mathcal{A}_{1}^{T} + \mathcal{A}_{1}S)\Gamma &\star &\star &\star \\
\mathcal{B}_{1}^{T}\Gamma & -\mu\mathbb{I}_{n_{e}} &\star &\star \\
G\mathcal{A}_{1}S\Gamma & 0 & -GSG^{T} &\star \\
\overline{C}_{p}S\Gamma & 0 & 0 & -\varepsilon\mathbb{I}_{n_{y}}
\end{array} \right) <  0
\end{array}
\end{equation}
and
\begin{equation}
  -\mathbb{I}_{n_{e}} < -GSG^{T}.
\end{equation}
As a consequence, it holds that
\begin{equation}\label{eq: prop-proof-ineq4}
\arraycolsep=1.4pt\def\arraystretch{1.5}
\left( \begin{array}{ccccccccc}
-\Gamma^{T}(S\mathcal{A}_{1}^{T} + \mathcal{A}_{1}S)\Gamma &\star &\star &\star \\
\mathcal{B}_{1}^{T}\Gamma & -\mu\mathbb{I}_{n_{e}} &\star &\star \\
G\mathcal{A}_{1}S\Gamma & 0 & -\mathbb{I}_{n_{e}} &\star \\
\overline{C}_{p}S\Gamma & 0 & 0 & -\varepsilon\mathbb{I}_{n_{y}}
\end{array} \right) < 0.
\end{equation}
Let $P = S^{-1}$ and pre and post multiply (\ref{eq: prop-proof-ineq4}) respectively by $\diag(P\Gamma^{-T}, \mathbb{I}_{n_{e}}, \mathbb{I}_{n_{e}}, \mathbb{I}_{n_{y}})$ and its transpose. Then, we have (note that $\mathcal{A}_{2} = G\mathcal{A}_{1}$)
\begin{equation}\label{eq: prop-proof-ineq5}
\begin{array}{c}
\left( \begin{array}{cccc}
{A}_{1}^{T}P + P\mathcal{A}_{1} & \star &\star &\star \\
\mathcal{B}_{1}^{T}P & -\mu\mathbb{I}_{n_{e}} &\star &\star \\
\mathcal{A}_{2} & 0 & -\mathbb{I}_{n_{e}} &\star \\
\overline{C}_{p} & 0 & 0 & -\varepsilon\mathbb{I}_{n_{y}}
\end{array} \right)  < 0.
\end{array}
\end{equation}
By using the Schur complement of (\ref{eq: prop-proof-ineq5}), we obtain
\begin{equation}\label{eq: proof-prop-LMI}
\left( \begin{array}{cccc} \mathcal{A}_{1}^{T}P + P\mathcal{A}_{1} + \mathcal{A}_{2}^{T}\mathcal{A}_{2} + \varepsilon_{1}\overline{C}_{p}^{T}\overline{C}_{p} & P\mathcal{B}_{1} \\ \mathcal{B}_{1}^{T}P & -\mu\mathbb{I}_{n_{e}} \end{array} \right) < 0,
\end{equation}
where $\varepsilon_{1} := \varepsilon^{-1}$. Hence, it holds that there exists $\varepsilon_{2} > 0$ sufficiently small such that
\begin{equation}\label{eq: proof-prop-LMI2}
\arraycolsep=1.4pt\def\arraystretch{1.2}
\left( \begin{array}{cccc} \mathcal{A}_{1}^{T}P + P\mathcal{A}_{1} + \mathcal{A}_{2}^{T}\mathcal{A}_{2} + \varepsilon_{1}\overline{C}_{p}^{T}\overline{C}_{p} + \varepsilon_{2}\mathbb{I}_{n_{x}} & P\mathcal{B}_{1} \\ \mathcal{B}_{1}^{T}P & -\mu\mathbb{I}_{n_{e}} \end{array} \right) \leq 0.
\end{equation}
Thus, Theorem \ref{thm: linear-systems-co-design} holds in virtue of Proposition 1 in \cite{Abdelrahim2014stabilization}. \hfill $\Box$

\vspace{0.5cm}
\noindent \textbf{Proof of Lemma \ref{lma-T-min}.}
Let $T^{*} := \underset{\phi\in\mathcal{S}}\inf\{t'-t: \exists j \in \Zg, \,\, (t,j), (t,j+1), (t',j+1), (t',j+2)\in \dom\phi\}$. The definitions of the flow and jump sets in (\ref{flow-jump-sets-out}) guarantee that $T^{*}\geq T$. We now show that $T^{*}\leq T$. Let $\widetilde{\phi}=(\widetilde{\phi}_{x},\widetilde{\phi}_{e},\widetilde{\phi}_{\tau})\in\mathcal{S}$ be such that $\widetilde{\phi}_{x}(0,0)=0, \widetilde{\phi}_{e}(0,0)=0, \widetilde{\phi}_{\tau}(0,0)=0$. Then, $\widetilde{\phi}_{x}(t,j)=0, \widetilde{\phi}_{e}(t,j)=0$ for all $(t,j)\in\dom\widetilde{\phi}$, in view of (\ref{eq: hybrid-model}). As a consequence, $\gamma^{2}|\widetilde{\phi}_{e}(t,j)|^{2}=\varepsilon_{1}|\widetilde{\phi}_{y}(t,j)|^{2}$ where $\widetilde{\phi}_{y}(t,j)=\overline{C}_{p}\widetilde{\phi}_{x}(t,j)$ for all $(t,j)\in\dom\widetilde{\phi}$ and two successive jumps are separated by $T$ units of time. We have that $T = \inf\{t'-t: \exists j \in \Zg, \,\, (t,j), (t,j+1), (t',j+1), (t',j+2)\in \dom\widetilde{\phi}\} \geq T^*$. Consequently $T = T^{*}$. \hfill $\Box$

\vspace{0.5cm}
\noindent \textbf{Proof of Claim \ref{clm: LMI3}.}
By using Schur complement of (\ref{eq: alpha-LMI}), we deduce that
\begin{equation}\label{eq: clm-proof-1}
\left(\begin{array}{ccccccccc}
 \alpha\mathbb{I}_{n_{y}}- Z^{T} Y^{-1} Z &\star &\star \\
0 & \beta\mathbb{I}_{n_{u}} &\star \\
- Y^{-1}  Z &  N^{T} &  X -  Y^{-1}
\end{array}\right) > 0.
\end{equation}
Re-applying the Schur complement of the last inequality yields
\begin{equation}\label{eq: clm-proof-aa}
\begin{array}{rll}
 X -  Y^{-1} &>& 0 \\
\left(\begin{array}{ccccccccc}
a_{11} & \star \\
a_{21} & a_{22}
\end{array}\right) &>& 0,
\end{array}
\end{equation}
where
\begin{equation}
\arraycolsep=1.4pt\def\arraystretch{1.2}
\begin{array}{rll}
a_{11} &:=&  \alpha\mathbb{I}_{n_{y}}- Z^{T} Y^{-1} Z- Z^{T} Y^{-1}( X -  Y^{-1})^{-1} Y^{-1} Z\\
a_{21} &:=&  N( X -  Y^{-1})^{-1} Y^{-1} Z \\
a_{22} &:=&  \beta\mathbb{I}_{n_{u}}-  N( X -  Y^{-1})^{-1} N^{T}.
\end{array}
\end{equation}
Using the fact that
\begin{equation}
  ( Y -  X^{-1})^{-1} =  Y^{-1} +  Y^{-1}( X -  Y^{-1})^{-1} Y^{-1}
\end{equation}
and since $( Y -  X^{-1})^{-1} > 0$ and $( X -  Y^{-1})^{-1}>0$, in view of the Schur complement of (\ref{eq: prop-ineq2}), inequality (\ref{eq: clm-proof-aa}) implies that
\begin{equation}
\left(\begin{array}{ccccccccc}
 \alpha\mathbb{I}_{n_{y}} & \star \\
 N( X -  Y^{-1})^{-1} Y^{-1} Z &  \beta\mathbb{I}_{n_{u}}
\end{array}\right) > 0.
\end{equation}
It holds that
\begin{equation}
\arraycolsep=1.4pt\def\arraystretch{1.2}
\begin{array}{rll}
  ( X -  Y^{-1})^{-1} Y^{-1} &=& ( Y( X -  Y^{-1}))^{-1} = ( Y X - \mathbb{I}_{n_{p}})^{-1} \\
  &=& - (\mathbb{I}_{n_{p}} -  Y X)^{-1}.
  \end{array}
\end{equation}
As a consequence
\begin{equation}
\left(\begin{array}{ccccccccc}
 \alpha\mathbb{I}_{n_{y}} & \star \\
- N(\mathbb{I}_{n_{p}} -  Y X)^{-1} Z &  \beta\mathbb{I}_{n_{u}},
\end{array}\right) > 0
\end{equation}
which implies that
\begin{equation}\label{eq: clm-proof-alphaI1}
   Z^{T}(\mathbb{I} -  Y X)^{-T} N^{T} N(\mathbb{I} -  Y X)^{-1} Z < \alpha\beta\mathbb{I}_{n_{y}}.
\end{equation}
On the other hand, in view (\ref{eq: prop-controller}), we have
\begin{equation}\label{eq: clm-proof-alphaI2}
\arraycolsep=1.4pt\def\arraystretch{1.2}
\begin{array}{rll}
  C_{c}B_{c} &=&  NU^{-T}V^{-1} Z =  N(UV^{T})^{-T} Z\\
             &=&  N(\mathbb{I}_{n_{p}} -  X Y)^{-T} Z =  N(\mathbb{I}_{n_{p}} -  Y X)^{-1} Z.
\end{array}
\end{equation}
As a result, in view of (\ref{eq: clm-proof-alphaI1}), (\ref{eq: clm-proof-alphaI2}), it holds that
\begin{equation}
   B_{c}^{T}C_{c}^{T}C_{c}B_{c}< \alpha\beta\mathbb{I}_{n_{y}}.
\end{equation}
Thus, Claim \ref{clm: LMI3} is verified. \hfill $\Box$

\bibliographystyle{IEEEtran}
\bibliography{D:/Mahmoud/References}
\end{document}